\documentclass[preprint]{ptephy}

\preprintnumber{KYUSHU-HET-151}

\usepackage{amssymb}
\usepackage{amsthm}
\usepackage{amsmath}
\usepackage{booktabs}
\usepackage{mathbbol}

\DeclareMathOperator{\Tr}{Tr}

\numberwithin{equation}{section}

\usepackage{url}
\let\Gamma\varGamma

\begin{document}

\title{Complex Langevin method applied to the 2D $SU(2)$ Yang-Mills theory}

\author{%
\name{\fname{Hiroki} \surname{Makino}}{1},
\name{\fname{Hiroshi} \surname{Suzuki}}{2,\ast}
\name{\fname{Daisuke} \surname{Takeda}}{3}
}

\address{%
\affil{1,2,3}{Department of Physics, Kyushu University, 6-10-1 Hakozaki,
Higashi-ku, Fukuoka, 812-8581, Japan}
\email{hsuzuki@phys.kyushu-u.ac.jp}
}

\begin{abstract}
The complex Langevin method in conjunction with the gauge cooling is applied to
the two-dimensional lattice $SU(2)$ Yang-Mills theory that is analytically
solvable. We obtain strong numerical evidence that at large Langevin time the
expectation value of the plaquette variable converges, but to a wrong value
when the complex phase of the gauge coupling is large.
\end{abstract}
\subjectindex{B01, B05, B34, B38}
\maketitle

\section{Introduction}
\label{sec:1}
As a possible approach to the functional integral with complex measure, such as
the one encountered in the finite density
QCD~\cite{deForcrand:2010ys,Gattringer:2014nxa}, the complex Langevin
method~\cite{Parisi:1984cs,Klauder:1983zm,Klauder:1983sp} has attracted much
attention in recent years. This recent interest was triggered mainly by the
discovery of sufficient conditions for the convergence of the method to a
correct answer~\cite{Aarts:2009uq,Aarts:2011ax}. Reference~\cite{Aarts:2013uxa}
is a review on recent developments. Roughly speaking, if the probability
distribution of configurations generated by the Langevin dynamics damps
sufficiently fast at infinity of configuration space, the statistical average
over the configurations is shown to be identical to the integration over the
original complex measure. It has been observed that, in systems for which the
complex Langevin (CL) method converges to a wrong answer (such as the
three-dimensional XY model~\cite{Aarts:2010aq}), this requirement of a
sufficiently localized distribution is broken, typically in ``imaginary
directions'' in configuration space.

After the above understanding, a prescription in lattice gauge theory that
makes the probability distribution well localized was proposed
in~Ref.~\cite{Seiler:2012wz}; the prescription is termed ``gauge cooling'' and
it proceeds as follows: The link variables in lattice gauge theory are
originally elements of the compact gauge group~$SU(N)$. When the (effective)
action is complex, however, the corresponding Langevin evolution drives link
variables into imaginary directions and link variables become elements
of~$SL(N,\mathbb{C})$, a \emph{noncompact\/} gauge group.\footnote{We will
shortly describe the Langevin evolution of link variables.} This evolution
tends to make the distribution wide in noncompact directions; in terms of the
$SU(N)$ Lie algebra, those noncompact directions are parametrized by imaginary
coordinates. At this point, one notes that the definition of a physical
observable that is invariant under the original $SU(N)$ gauge transformations
can always be tailored so that it is invariant also under the noncompact
$SL(N,\mathbb{C})$ gauge transformations. The idea of the gauge cooling is that
by applying the $SL(N,\mathbb{C})$ gauge transformations appropriately along
the complex Langevin evolution, one squeezes the distribution well localized so
that the prerequisite of the convergence
theorem~\cite{Aarts:2009uq,Aarts:2011ax} is fulfilled without changing physical
observables.

In a one-dimensional gauge model and in the four-dimensional QCD with heavy
quarks, it has been confirmed that the gauge cooling makes the distribution
well localized and the complex Langevin method gives rise to correct
answers~\cite{Seiler:2012wz}. More recently, this method was applied to the
full QCD at finite density~\cite{Sexty:2013ica}. See also
Refs.~\cite{Mollgaard:2013qra,Splittorff:2014zca,Mollgaard:2014mga}. One should
note, however, that the Langevin dynamics itself is defined on gauge
noninvariant variables (i.e., link variables) and also that the gauge cooling
step cannot be regarded as a Langevin evolution that is induced by a
\emph{holomorphic\/} action; the latter is assumed in the convergence
theorem~\cite{Aarts:2009uq,Aarts:2011ax}. Strictly speaking, therefore, the
convergence theorem does not apply when the gauge cooling is employed. The
method should still be carefully examined in various possible ways.

In the present paper, we apply the complex Langevin method in conjunction with
the gauge cooling to the two-dimensional lattice Yang-Mills theory which can be
analytically solved~\cite{Balian:1974ts,Migdal:1975zg,Rothe:1992nt}. By doing
this, we examine the validity of the method. The partition function of the
two-dimensional Yang-Mills theory on the lattice is given by
\begin{equation}
   \mathcal{Z}=\int\left[\prod_{x,\mu}\mathrm{d}U_{x,\mu}\right]
   \mathrm{e}^{-S},
\label{eq:(1.1)}
\end{equation}
where $U_{x,\mu}$ are link variables defined on a two-dimensional rectangular
lattice, $S$ is the lattice action,\footnote{Throughout this paper, $N=2$.}
\begin{equation}
   S=-\frac{\beta}{2N}\sum_x
   \Tr\left[U_{01}(x)+U_{01}(x)^{-1}\right],
\label{eq:(1.2)}
\end{equation}
and the plaquette variable is defined by
\begin{equation}
   U_{\mu\nu}(x)
   =U_{x,\mu}U_{x+\Hat{\mu},\nu}
   U_{x+\Hat{\nu},\mu}^{-1}U_{x,\nu}^{-1}.
\label{eq:(1.3)}
\end{equation}

For simplicity, we assume that the gauge group is~$SU(2)$, that is,
$U_{x,\mu}\in SU(2)$ in the original integral~\eqref{eq:(1.1)}. On the other
hand, when the gauge coupling~$\beta$ is complex, the corresponding Langevin
equation [Eq.~\eqref{eq:(2.1)} below] evolves link variables as elements
of~$SL(2,\mathbb{C})$. Thus, the distinction between $U_{x,\mu}^\dagger$
and~$U_{x,\mu}^{-1}$ becomes very important in the complex Langevin dynamics.
For the convergence theorem in~Refs.~\cite{Aarts:2009uq,Aarts:2011ax} to apply,
the action~$S$ that generates the drift force in the Langevin equation and
physical observables must be a holomorphic function of dynamical variables; our
above definitions~\eqref{eq:(1.2)}--\eqref{eq:(1.3)} that entirely use
$U_{x,\mu}^{-1}$ not~$U_{x,\mu}^\dagger$ are chosen by this criterion. Note also
that the plaquette action~\eqref{eq:(1.2)} is invariant under the
$SL(2,\mathbb{C})$ lattice gauge transformations [such as the one
in~Eq.~\eqref{eq:(2.4)}].

We consider the expectation value of the plaquette variable:
\begin{equation}
   \left\langle\Tr\left[U_{01}(x)\right]\right\rangle
   =\frac{1}{\mathcal{Z}}
   \int\left[\prod_{x,\mu}\mathrm{d}U_{x,\mu}\right]
   \mathrm{e}^{-S}
   \Tr\left[U_{01}(x)\right].
\label{eq:(1.4)}
\end{equation}
Even if the gauge coupling~$\beta$ is complex, this can be exactly computed by
the character expansion~\cite{Balian:1974ts,Migdal:1975zg,Rothe:1992nt}. Under
periodic boundary conditions, one yields
\begin{equation}
   \left\langle\Tr\left[U_{01}(x)\right]\right\rangle
   =-\frac{N}{V}\frac{\partial}{\partial\beta}\ln\mathcal{Z},\qquad
   \mathcal{Z}=\sum_{n=1}^\infty\left[\frac{2}{\beta}I_n(\beta)\right]^V,
\label{eq:(1.5)}
\end{equation}
where $I_n(x)$ denotes the modified Bessel function of the first kind and
$V$~is the number of lattice points.

\section{Complex Langevin method and the gauge cooling}
\label{sec:2}
The following procedures are basically identical to the ones adopted for the
four-dimensional lattice QCD in~Ref.~\cite{Sexty:2013ica} for example, although
our two-dimensional pure-gauge system is much simpler.

For the link variable, the Langevin equation with a discretized Langevin
time~$t$ with the time step~$\epsilon$ is defined by
\begin{equation}
   U_{x,\mu}(t+\epsilon)
   =\exp\left[i\sum_a\lambda_a
   \left(\sqrt{\epsilon}\eta_{a,x,\mu}(t)-\epsilon D_{a,x,\mu}S\right)
   \right]U_{x,\mu}(t),
\label{eq:(2.1)}
\end{equation}
where $\lambda_a$ ($a=1$, $2$, $3$) are Pauli matrices, $\eta_{a,x,\mu}(t)$ are
Gaussian real random numbers of the variant
\begin{equation}
   \left\langle\eta_{a,x,\mu}(t)\eta_{b,y,\nu}(t')\right\rangle
   =2\delta_{ab}\delta_{xy}\delta_{\mu\nu}\delta_{tt'},
\label{eq:(2.2)}
\end{equation}
and $D_{a,x,\mu}S$ is the drift force generated by the action~$S$
in~Eq.~\eqref{eq:(1.2)}; the derivative with respect to the link variable is
given by
\begin{equation}
   D_{a,x,\mu}f(U)=\left.\partial_\xi
   f(\mathrm{e}^{i\xi\lambda_a}U_{x,\mu})\right|_{\xi=0}.
\label{eq:(2.3)}
\end{equation}
When the gauge coupling~$\beta$ is complex, the drift force becomes complex and
the Langevin evolution evolves link variables as elements
of~$SL(2,\mathbb{C})$.

The above complex Langevin dynamics tends to make the probability distribution
function of link variables wide in noncompact directions of~$SL(2,\mathbb{C})$.
To squeeze the distribution well localized without changing gauge invariant
quantities, we apply the following $SL(2,\mathbb{C})$ gauge transformation
(this step is the gauge cooling)
\begin{equation}
   U_{x,\mu}\to U_{x,\mu}'=V_xU_{x,\mu}V_{x+\Hat{\mu}}^{-1},
\label{eq:(2.4)}
\end{equation}
where
\begin{equation}
   V_x=\mathrm{e}^{-\epsilon\alpha f_a^x\lambda_a},\qquad
   f_a^x=2\Tr\left[\lambda_a\sum_\mu
   \left(
   U_{x,\mu}U_{x,\mu}^\dagger-U_{x-\Hat{\mu},\mu}^\dagger U_{x-\Hat{\mu},\mu}
   \right)\right],
\label{eq:(2.5)}
\end{equation}
and $\alpha>0$ is a real parameter. The distance defined by~\cite{Aarts:2008rr}
\begin{equation}
   d=\frac{1}{V}\sum_{x,\mu}
   \frac{1}{N}\Tr\left(U_{x,\mu}U_{x,\mu}^\dagger-\mathbb{1}\right)\geq0
\label{eq:(2.6)}
\end{equation}
measures how a $SL(2,\mathbb{C})$ gauge field is far away from the subspace of
$SU(2)$ gauge fields. It is then straightforward to see that for a sufficiently
small~$\epsilon$, the gauge cooling~\eqref{eq:(2.4)} decreases or does not
change the distance~$d$. Note that $f_a^x$ in~Eq.~\eqref{eq:(2.5)} is not a
holomorphic function of link variables and thus the step~\eqref{eq:(2.4)}
cannot be regarded as a part of the complex Langevin dynamics in which the
drift force is generated by a holomorphic action; this fact prevents us from
applying the convergence theorem~\cite{Aarts:2009uq,Aarts:2011ax} to the above
procedures.

\section{Result of numerical simulations}
\label{sec:3}
We numerically solved the Langevin equation with the discretized Langevin time,
Eq.~\eqref{eq:(2.1)}, on a $V=4^2$ lattice. Periodic boundary conditions are
imposed. The maximal size of the time step~$\epsilon$ we adopted was~$0.001$
and, when the drift force becomes large, we further reduced $\epsilon$
``adaptively'' according to the prescription in~Ref.~\cite{Aarts:2009dg}.

As the parameter~$\alpha$ in the gauge cooling~\eqref{eq:(2.5)}, we tried both
$\alpha=1$ and the adaptive choice (see Ref.~\cite{Aarts:2013uxa})
\begin{equation}
   \alpha_{\text{ad}}=\frac{1}{D},\qquad
   D\equiv\frac{1}{V}\sum_{a,x}|f_a^x|+1.
\label{eq:(3.1)}
\end{equation}
Our numerical results did not show any notable difference in these two choices
and we present the results with the latter choice in what follows.

\begin{figure}[ht]
\begin{center}
\includegraphics[scale=0.8,clip]{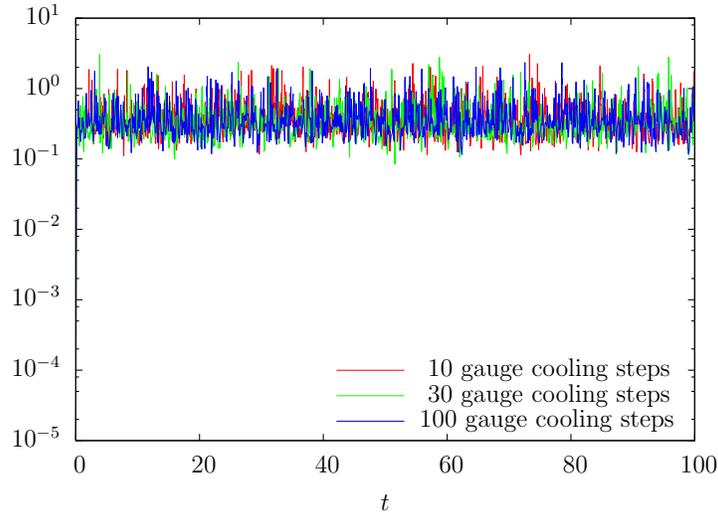}
\caption{Evolution of the distance~$d$~\eqref{eq:(2.6)} for~$\beta=0.4+2.0\,i$
with various numbers of the gauge cooling steps per one Langevin
update~\eqref{eq:(2.1)}, $10$, $30$, and~$100$.}
\label{fig:1}
\end{center}
\end{figure}

\begin{figure}[ht]
\begin{center}
\includegraphics[scale=0.8,clip]{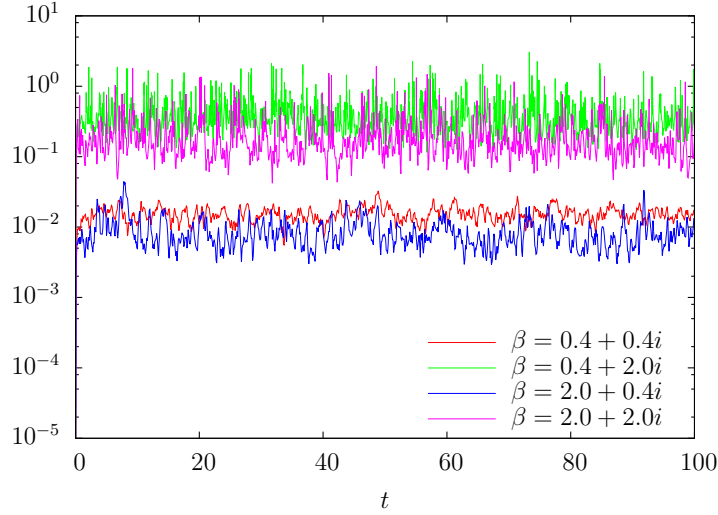}
\caption{Evolution of the distance~$d$~\eqref{eq:(2.6)} for various complex
gauge couplings, $\beta=0.4+0.4\,i$, $0.4+2.0\,i$, $2.0+0.4\,i$,
and~$2.0+2.0\,i$. The number of the gauge cooling steps per one Langevin
update~\eqref{eq:(2.1)} is~$10$.}
\label{fig:2}
\end{center}
\end{figure}

To determine an appropriate rate of the gauge cooling~\eqref{eq:(2.4)} along
the Langevin evolution, we observed the time evolution of the
distance~$d$~\eqref{eq:(2.6)} by changing the number of the gauge cooling steps
per one Langevin update~\eqref{eq:(2.1)}. In~Fig.~\ref{fig:1},
for~$\beta=0.4+2.0\,i$, we plotted the evolution of the distance~$d$ as the
function of the Langevin time~$t$ by changing the number of the gauge cooling
steps per one Langevin update as, $10$, $30$, and~$100$.\footnote{We define the
Langevin time~$t$ such that it does not elapse during the gauge cooling steps.}
Since this plot shows the evolution including the Langevin stochastic dynamics,
the distance~$d$ does not necessarily decrease.
Since we do not see much difference for those three choices, we adopted ten
gauge cooling steps per one Langevin update. With this choice, the evolution of
the distance~$d$ for various complex gauge couplings, $\beta=0.4+0.4\,i$,
$0.4+2.0\,i$, $2.0+0.4\,i$, and~$2.0+2.0\,i$, looks as depicted
in~Fig.~\ref{fig:2}.
It appears that the gauge cooling is working perfectly for those complex gauge
couplings, suppressing the evolution to noncompact imaginary directions.

\begin{figure}[ht]
\begin{center}
\includegraphics[scale=0.8,clip]{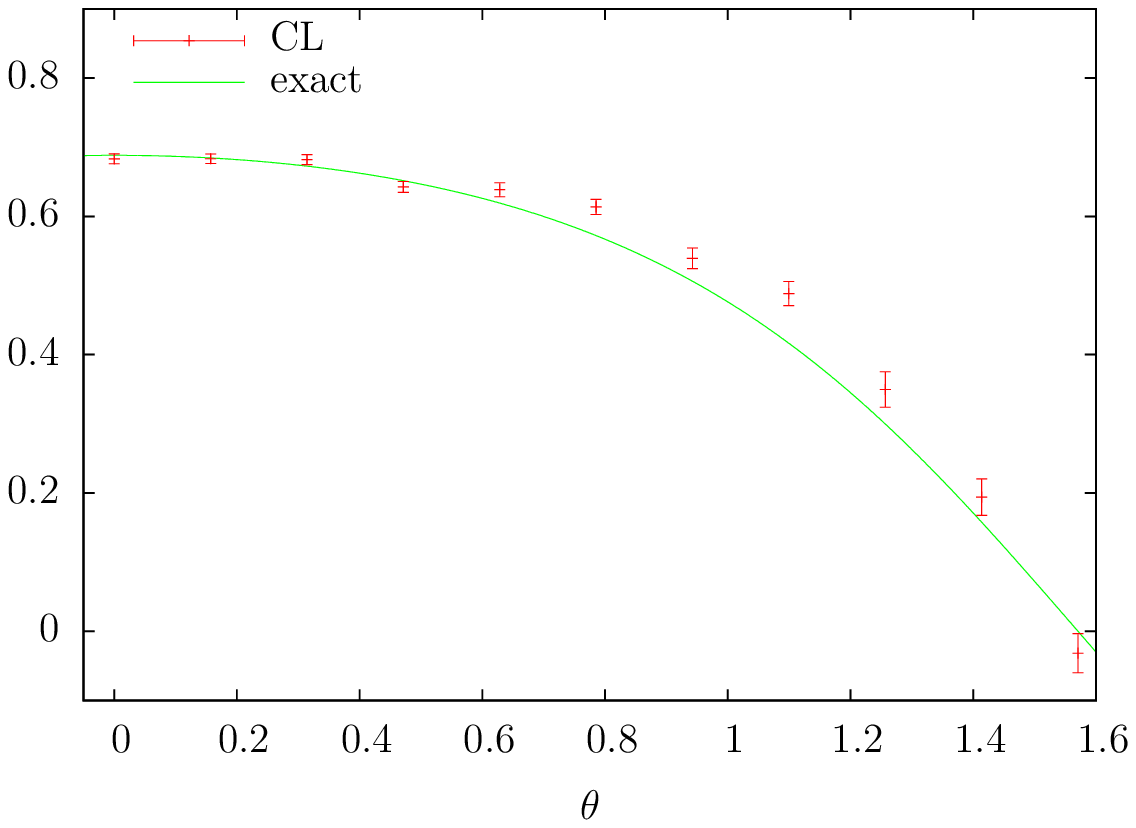}
\caption{Real part of the expectation value~\eqref{eq:(1.4)} obtained by the CL
method and the exact value given by~Eq.~\eqref{eq:(1.5)}. The horizontal axis
is the complex phase~$\theta$ of the gauge coupling in~Eq.~\eqref{eq:(3.2)}.}
\label{fig:3}
\end{center}
\end{figure}

\begin{figure}[ht]
\begin{center}
\includegraphics[scale=0.8,clip]{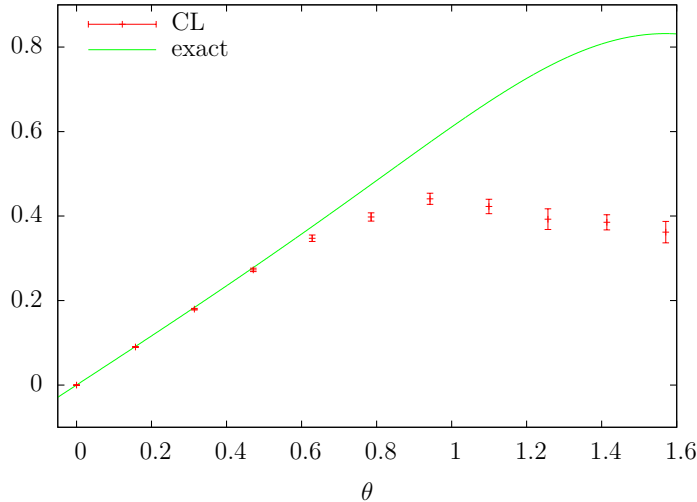}
\caption{Imaginary part of the expectation value~\eqref{eq:(1.4)} obtained by
the CL method and the exact value given by~Eq.~\eqref{eq:(1.5)}. The horizontal
axis is the complex phase~$\theta$ of the gauge coupling
in~Eq.~\eqref{eq:(3.2)}.}
\label{fig:4}
\end{center}
\end{figure}

Now, we turn to the computation of the expectation value of the plaquette,
Eq.~\eqref{eq:(1.4)}, by the complex Langevin method. Starting from a
configuration of random $SU(2)$ matrices, we discarded configurations until the
Langevin time~$t=11$ for thermalization. Then $1000$ configurations separated
by~$\Delta t=1$ from $t=11$ to~$t=1010$ are used to compute the expectation
value. For typical values of the complex gauge coupling, we confirmed that the
plaquette values between configurations separated by~$\Delta t=1$ practically
have no autocorrelation. In Figs.~\ref{fig:3}--\ref{fig:4}, we plotted the real
and imaginary parts of the expectation value~\eqref{eq:(1.4)} obtained by the
CL method. The error bars are statistical ones. The horizontal axis is the the
complex phase~$\theta$ of the gauge coupling\footnote{When the gauge group
is~$SU(2)$, the partition function~\eqref{eq:(1.1)} is invariant
under~$\beta\to-\beta$. Because of this invariance and the complex conjugation,
it is sufficient to consider the range of the phase, $0\leq\theta\leq\pi/2$.}
with the modulus~$1.5$:
\begin{equation}
   \beta=1.5\,\mathrm{e}^{i\theta},\qquad0\leq\theta\leq\pi/2.
\label{eq:(3.2)}
\end{equation}
The solid line curves are exact values given by~Eq.~\eqref{eq:(1.5)}. We see
that the complex Langevin method reproduces the real part fairly well, while it
clearly fails to converge to the correct value of the imaginary part when the
complex phase of the gauge coupling is large.

\begin{figure}[ht]
\begin{center}
\includegraphics[scale=0.8,clip]{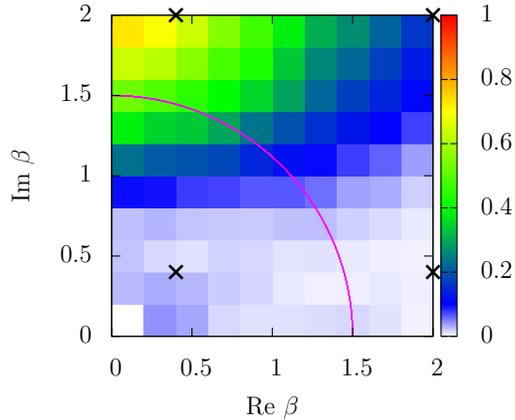}
\caption{Gradation plot of the relative error~\eqref{eq:(3.3)} on the complex
$\beta$ (the gauge coupling) plane. The block around the origin~$\beta=0$ is
omitted from the plot. The quadrant is~Eq.~\eqref{eq:(3.2)} along which
Figs.~\ref{fig:3}--\ref{fig:4} are plotted. Four black crosses indicate complex
gauge couplings we used in~Fig.~\ref{fig:2}.}
\label{fig:5}
\end{center}
\end{figure}

The gradation plot in~Fig.~\ref{fig:5} shows the relative error
\begin{equation}
   \frac
   {\left|
   \left\langle\Tr\left[U_{01}(x)\right]\right\rangle_{\text{CL}}
   -\left\langle\Tr\left[U_{01}(x)\right]\right\rangle_{\text{exact}}
   \right|}
   {\left|
   \left\langle\Tr\left[U_{01}(x)\right]
   \right\rangle_{\text{exact}}
   \right|}
\label{eq:(3.3)}
\end{equation}
on the first quadrant of the complex~$\beta$ plane.\footnote{The block around
the origin~$\beta=0$ is omitted from the plot because
$\langle\Tr[U_{01}(x)]\rangle\sim0$ for~$\beta\sim0$.}
The (quadrant) circle in the figure is~Eq.~\eqref{eq:(3.2)} along which
Figs.~\ref{fig:3}--\ref{fig:4} are plotted. Clearly, the relative error of the
complex Langevin method becomes large when the complex phase of the gauge
coupling becomes large. Four black crosses in the figure indicate complex gauge
couplings we used in~Fig.~\ref{fig:2}; the behavior in~Fig.~\ref{fig:2} thus
suggests that the gauge cooling is correctly working for the region of the
complex gauge coupling shown in~Fig.~\ref{fig:5}. Nevertheless, the complex
Langevin method shows large deviation from the correct value as
in~Fig.~\ref{fig:4}. This is the main result of the present paper.

A similar failure of the complex Langevin method for large complex $\beta$ has
been observed; see Fig.~8 of~Ref.~\cite{Aarts:2012ft}. This result
of~Ref.~\cite{Aarts:2012ft} is, however, for a one-dimensional integral (not a
gauge theory) and the validity of the gauge cooling, which is our main issue in
this paper, is not relevant to this result of~Ref.~\cite{Aarts:2012ft}.

\begin{figure}[ht]
\begin{center}
\includegraphics[scale=0.8,clip]{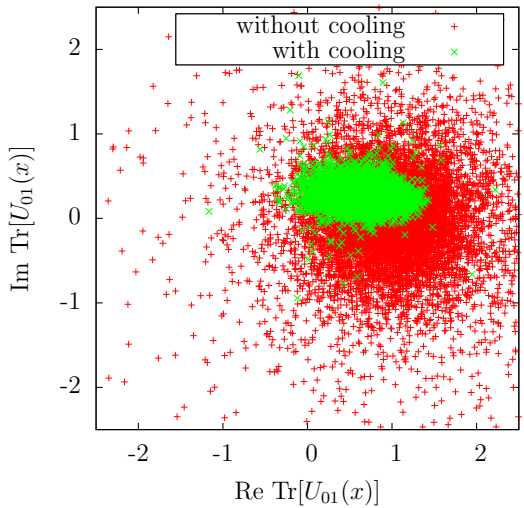}
\caption{Distribution of the plaquette variable averaged over the lattice
volume. $\beta=1.5\,\mathrm{e}^{i\,(0.3\pi/2)}$ ($\theta=0.3\pi/2$).}
\label{fig:6}
\end{center}
\end{figure}

\begin{figure}[ht]
\begin{center}
\includegraphics[scale=0.8,clip]{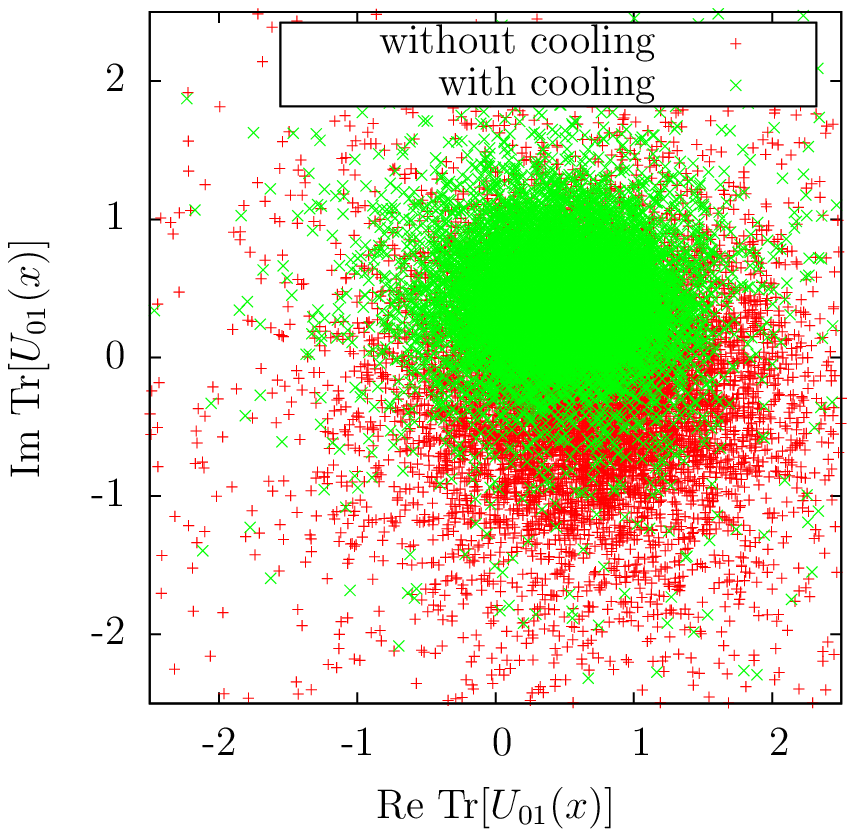}
\caption{Distribution of the plaquette variable averaged over the lattice
volume. $\beta=1.5\,\mathrm{e}^{i\,(0.7\pi/2)}$ ($\theta=0.7\pi/2$).}
\label{fig:7}
\end{center}
\end{figure}

It is of interest how configurations generated by the Langevin dynamics
distribute in configuration space. To give some idea on this point,
in~Figs.~\ref{fig:6}--\ref{fig:7}, we present scatter plots of the plaquette
variable (averaged over the lattice volume) for each configuration.
Both cases, with and without the gauge cooling, are shown. Figure~\ref{fig:6}
is for~$\beta=1.5\,\mathrm{e}^{i\,(0.3\pi/2)}$ (i.e., $\theta=0.3\pi/2$) and
corresponds to points in~Figs.~\ref{fig:3}--\ref{fig:4} with a relatively small
complex phase. Figure~\ref{fig:7} is, on the other hand,
for~$\beta=1.5\,\mathrm{e}^{i\,(0.7\pi/2)}$ (i.e., $\theta=0.7\pi/2$) and
corresponds to points in~Figs.~\ref{fig:3}--\ref{fig:4} with a large complex
phase and with large deviation. Although there is a tendency when the complex
phase of the gauge coupling is large for the distribution to become somewhat
wider even after the gauge cooling, it is not clear from the scatter plot
in~Fig.~\ref{fig:4} alone whether the distribution is so poorly localized as to
break the prerequisite of the convergence
theorem~\cite{Aarts:2009uq,Aarts:2011ax}. More detailed study is needed on this
point.

\section{Conclusion}
\label{sec:4}
In the present paper, we applied the complex Langevin method in conjunction
with the gauge cooling to the two-dimensional lattice $SU(2)$ Yang-Mills
theory. Our intention was to examine the validity of the method by using this
analytically solvable model. Somewhat unexpectedly, as shown
in~Figs.~\ref{fig:4}--\ref{fig:5}, we obtained strong numerical evidence that
the method fails to converge to the correct value when the complex phase of the
gauge coupling is large. As we emphasized in the introduction, the convergence
proof of~Refs.~\cite{Aarts:2009uq,Aarts:2011ax} does not necessarily apply when
the gauge cooling is employed; thus there is no contradiction even if the
method leads to a wrong answer. Nevertheless, it is not yet clear what causes
the failure for the gauge coupling with a large complex phase. To find the
resolution of the problem we found in the present study, first we have to pin
down what the real source of the failure is. For this, consideration on the
basis of another approach to the functional integral with complex measure, the
Lefschetz thimble~\cite{Cristoforetti:2012su,Cristoforetti:2013wha,%
Fujii:2013sra,Aarts:2014nxa,Kanazawa:2014qma}, might provide useful insight.
See also Ref.~\cite{Cherman:2014xia} for suggestive observations.

\section*{Acknowledgements}

We thank Erhard Seiler, Kim Splittorff, and Ion-Olimpiu Stamatescu for the
introduction to the complex Langevin method. We are grateful to Masanobu Yahiro
for encouragement. The work of H.~S. is supported in part by Grant-in-Aid for
Scientific Research Grant No.~23540330.

\end{document}